\documentclass[]{Liebert_Author}

\usepackage{authblk}
\usepackage{natbib}
\usepackage{graphicx}% Include figure files
\usepackage[version=4]{mhchem}
\usepackage[english]{babel}
\usepackage[utf8]{inputenc}
\usepackage[margin=1in]{geometry}
\usepackage{amsmath}
\usepackage{csquotes}% Recommended

\def\E{\mathrm{e}}
\def\perK{~K$^{-1}$}
\def\Wm2K{~W~m$^{-2}$\perK}
\def\Wmsq{W~m$^{-2}$}

\title{Self-sustaining living habitats in extraterrestrial environments} 

\author[1,2,*]{R. Wordsworth}
\author[3]{C. Cockell}
\affil[1]{School of Engineering and Applied Sciences, Harvard, Cambridge, MA 02138, USA}
\affil[2]{Department of Earth and Planetary Sciences, Harvard, Cambridge, MA 02138, USA}
\affil[3]{School of Physics and Astronomy, University of Edinburgh, Scotland, UK}
\date{}

\begin{document} 

\maketitle 

\begin{abstract}
Standard definitions of habitability assume that life requires the presence of planetary gravity wells to stabilize liquid water and regulate surface temperature. Here the consequences of relaxing this assumption are evaluated. Temperature, pressure, volatile loss, radiation levels and nutrient availability all appear to be surmountable obstacles to the survival of photosynthetic life in space or on celestial bodies with thin atmospheres. Biologically generated barriers capable of transmitting visible radiation, blocking {ultraviolet}, and sustaining temperature gradients of 25-100 K and pressure differences of 10 kPa against the vacuum of space can allow habitable conditions between 1 and 5 {astronomical units} in the solar system. Hence ecosystems capable of generating conditions for their own survival are physically plausible, given the known capabilities of biological materials on Earth. Biogenic habitats for photosynthetic life in extraterrestrial environments would have major benefits for human life support and sustainability in space. Because the evolution of life elsewhere may have followed very different pathways from on Earth, living habitats could also exist outside traditional habitable environments around other stars, where they would have unusual but potentially detectable biosignatures.
\end{abstract}

\section{Introduction}

Earth is currently our only known example of a planet that sustains life. As a result, a common working assumption in astrobiology is that Earth-like environments are required for life to exist anywhere. Exceptions to this rule do exist, in the form of speculations about exotic life on gas giants, Titan and Venus, for example \citep{sagan1976particles,mckay2005possibilities,seager2021venusian}. However, in biosignature definitions and mission planning, it is often argued that a focus on conservative definitions of habitability is most appropriate \citep{kasting2014remote}. In keeping with this view, most astrobiology research over the past few decades has focused on increasing our understanding of life's diversity and evolution on Earth, and searching for Earth-like environments (past or present) in the solar system and beyond.

One obvious yet often neglected fact in this debate is that we already have direct evidence of life existing beyond Earth, in the form of human space missions. Given current developments, it seems reasonable to assume that within the next few decades, the permanent presence of human life beyond low Earth orbit will be possible. Of course, this presence will be sustained by technology, which is ultimately a product of human intelligence. However, humans are simply a particularly complex form of life, so it is interesting to consider \emph{how much} complexity is really needed for life to sustain itself beyond Earth. Put another way, what is the minimal physical structure that could sustain habitable conditions beyond Earth, and could it feasibly be generated by non-sentient living organisms? 

Here we suggest that this question is both worthy of serious investigation and experimentally testable. {We} describe how the key physical challenges to life in space can be overcome {and} propose that further research on this topic will have major benefits for both astrobiology and the {study of human life support} \citep{nangle2020case,berliner2021towards}. We mainly consider carbon-based life, but wider possibilities are also briefly discussed. Our work draws inspiration from previous speculation on how living organisms might survive in space, dating back as far as the 19\textsuperscript{th} century \citep{tsiolkovsky1895speculation,dyson1979disturbing,sagan1997comet,church2022picogram}. 

We begin by defining the physical requirements for life beyond Earth, and then show quantitatively how they can be met by known biological materials.  We then discuss several key features that a self-sustaining extraterrestrial ecosystem might exhibit. Our focus here is on habitability for photosynthetic life rather than humans, although our conclusions have implications for human life support. We also mostly consider habitability in deep space, although our discussion also applies to more benign environments such as the martian surface. 

In Section~2, we briefly review key concepts in the habitability of planets and icy moons. In Sections 3 to 6, we discuss how habitability challenges {including} temperature, pressure {and} volatile retention can be overcome. In Section 7, we discuss how these physical constraints help determine the optimal scale and location of living habitats. In Sections 8 and 9 we briefly discuss questions related to maintenance, growth and non carbon-based life. In Section 10 we conclude and discuss future implications.

\section{Planetary Habitability}

To understand the constraints on life beyond Earth, we can start by reviewing why our home planet is a good habitat for life in the first place. 
Earth has a well-defined surface at which temperature and pressure are in the right range to allow the vital solvent \ce{H2O} to exist in liquid form. At this surface, a huge source of free energy is nonetheless available in the form of the Sun's photons, which our solar-powered biosphere exploits via photosynthesis. The essential elements C, H, N, O, P and S required to construct biochemical machinery are often limited, but nevertheless available to life. When the biosphere converts them into inaccessible waste products such as recalcitrant organic matter, they are slowly recycled back into more accessible forms via plate tectonics and volcanic outgassing. Finally, Earth is oxidizing in some regions (the atmosphere, surface environments) and reducing in others (deep subsurface, locally in surface soils, ocean sediments), allowing the exploitation of redox gradients for metabolic purposes. These are the key features of our planet that have allowed it to remain habitable over the last four billion years.

On smaller planets, habitable environments can still exist, but problems begin to emerge as the planet ages. First, internal cooling proceeds more quickly. {This} leads to an early shutdown in tectonic activity, as happened on Mars in the first billion years of its history. Second, surface gravity {is lower}, which makes it easier for any volatiles to escape to space. Again, Mars is {a case study} for volatile loss in our solar system: isotopic analysis tells us that the vast majority of its \ce{CO2} atmosphere, as well as much of its surface water inventory, has been lost to space over its lifetime \citep{jakosky2021atmospheric}. 

Further out in the solar system, many icy moons and dwarf planets are rich in life-supporting elements, but their surfaces are far too cold to allow liquid water. Moons such as Europa that experience tidal heating have subsurface oceans that likely support habitable conditions \citep{carr1998evidence}. However, tens of kilometers of ice separate these regions from the Sun's photons, making solar radiation inaccessible as a direct energy source for photosynthesis. At the end of the Sun's evolution on the main sequence, icy moons like Europa may pass through a transient stage where surface liquid water is possible, but their atmospheres will be highly unstable to loss to space \citep{arnscheidt2019atmospheric}. At still lower masses, volatile-rich icy bodies become comets: if they get close to the Sun, they degas directly to space without the water on their surface ever passing through a liquid phase at all. 

For low-mass objects outside of planetary gravity wells, then, pressure, temperature and long-term volatile retention are all fundamental challenges to habitability for Earth-like life \citep{cockell2021minimum}. Additional challenges include {ultraviolet (UV)} and cosmic ray radiation, availability of nutrients, and chemical toxicity. To persist beyond Earth, any living organism must modify or adapt to its environment enough to surmount these challenges. In the following sections, we evaluate each of these challenges in turn and show how they can be overcome in practice {using biologically generated materials}.

\section{Pressure}\label{sec:pres}

{Habitats}\footnote{{A note on terminology: in general, `habitat' is any physical region within which life can be sustained. Here, we are considering habitats that are biologically generated structures. For photosynthetic carbon-based life, `habitat wall' refers to a boundary that permits the transmission of visible-wavelength photons but inhibits incoming UV radiation, outgoing infrared, and loss of volatiles.}}
 surrounded by a vacuum or low pressure atmosphere must maintain a pressure gradient {across their walls} if they are to stabilize liquid water {internally}. The minimum pressure required to sustain liquid water is the triple point: 611.6~Pa at $0^\circ$C (273~K). {In} the $15-25^\circ$C temperature range, {this minimum} pressure rises to a few kPa. For comparison, sea level pressure on Earth is 101~kPa, while the atmospheric pressure at the summit of Mount Everest is about 30~kPa. {A few kPa is close to the experimentally determined limits for growth of simple organisms.}
For example, cyanobacteria in a liquid medium will grow in chambers with air headspace pressures of 10 kPa, provided that their nutrient needs are met and light and pH levels are within an acceptable range \citep{verseux2020bacterial}.

Internal pressure differences of order 10 kPa {are easily maintained by biological materials and in fact} common in macroscopic organisms on Earth. Using the hydrostatic equation $p = \rho gh$, it can be seen that the blood pressure increase from the head to the feet of a 1.5-m tall human is around 15~kPa, given gravity $g = 9.8$~m~s$^{-2}$ and blood density $\rho \sim 1000$~kg~m$^{-3}$. In the tallest land mammals, this rises to about 50~kPa. In plants, the seaweed species \emph{Ascophyllum nodosum} has been reported to sustain internal float nodule pressures of 15-25~kPa due to daytime \ce{O2} gas release from photosynthesis \citep{damant1937storage}. 

{We can calculate the habitat wall thickness required to sustain a given pressure difference using the hoop stress equation}. Given an idealized thin-walled spherical habitat, the hoop stress can be written as
\begin{equation}
\sigma_\theta = \frac{pr}{2t},
\end{equation}
where $p$ is internal pressure, $r$ is the habitat radius and $t$ is the wall thickness \citep{goodno2020mechanics}. If $r=1$~m, $p=10$~kPa and $t=0.05$~m, {then} $\sigma_\theta = 0.1$~MPa. For comparison, translucent bioplastics based on polysaccharides such as agarose have tensile strengths in the 10-40~MPa range \citep{hernandez2022agar}. Integration of silica or calcium carbonates into wall material would increase robustness further, particularly when combined with the ability of biology to optimize material properties down to nanometer scales. For example, the calcite shells of windowpane oysters (\emph{Placuna Placenta}) are near-transparent (up to $\sim$80\% transmittance of visible light given thicknesses of a few mm) but have almost order-of-magnitude greater resistance to localized fracture damage than crystalline calcite minerals \citep{li2014pervasive}. Overall, therefore, {pressure regulation does not} present a serious impediment to habitability.

\section{Temperature}\label{sec:temp}

Earth today is kept warm {enough for liquid water} by an atmospheric greenhouse effect: incoming sunlight is absorbed by the surface but upwelling infrared radiation cannot escape to space until it reaches high, cold regions of the atmosphere, which leads to a net warming of the surface \citep{pierrehumbert2010principles}. In space, retaining an atmosphere is impossible for low-mass objects. {Hence a biologically generated habitat must achieve} the same effect via solid-state physics.

For a free-floating object in a vacuum, the internal temperature in equilibrium can be calculated by equating the absorbed solar radiation to the emitted infrared radiation. The absorbed solar radiation is $A_1 F_0 \alpha / d^2$, where $A_1$ is the cross-sectional area of the object facing the Sun, $\alpha$ is visible absorptivity, $F_0 = 1362$~W~m$^{-2}$ is the solar flux at 1~AU and $d$ is orbital distance in AU. The emitted infrared radiation is $A_2 \epsilon \sigma T^4$, where $A_2$ is the emitting area, $\epsilon$ is {infrared} emissivity and $\sigma$ is Stefan's constant. Equating these quantities and writing equilibrium temperature $T_{eq}=T$ leads to the expression
\begin{equation}
T_{eq} = \left(\frac{f F_0}{\sigma d^2} \frac \alpha {\epsilon} \right)^{1/4}.\label{eq:Teq}
\end{equation}
The geometric factor $f\equiv A_1/A_2$ takes the value 0.25 for a sphere, and 0.5 for a flattened disk. On Earth today, changes in both $\alpha$ and $\epsilon$ are already utilized by living organisms to regulate temperature. Saharan silver ants, for example, have evolved the ability to enhance both their surface near-infrared reflectivity and their thermal emissivity, allowing them to survive in ambient temperatures above the range of all other known arthopods \citep{shi2015keeping}. 

{Besides varying} surface absorptivity and emissivity, the {solid-state greenhouse effect} provides a powerful means to regulate habitat temperature. This effect arises in translucent solid layers that transmit visible radiation but limit thermal radiation and conduction. The solid-state greenhouse can be potent under the right conditions: silica aerogels, for example, have been shown to raise temperatures by 50~K over 2-3 cm layers at 1~bar pressure, given a visible flux of 150~W~m$^{-2}$ \citep{wordsworth2019enabling}. In a vacuum, aerogel warming potential is higher still, because there is no longer a contribution to thermal conductivity from interstitial air molecules \citep{dorcheh2008silica}. 

Silica aerogels are industrially produced materials consisting of interlocking nanoscale \ce{SiO2} networks that have no direct biological equivalent. However, many organisms do exist in nature that produce complex silica structures. Diatoms, for example, have the ability to manipulate silica particles on 1-10~nm scales, which is below the mean pore scale for modern manufactured silica aerogels. {Furthermore,} {recent research has shown that aerogels manufactured directly from} organic materials such as cellulose have similar thermal properties \citep{smalyukh2021thermal}. {Given this, {it is plausible that} highly insulating materials {could be produced artificially} from biogenic feedstocks, or even directly by living organisms.

\begin{figure}[h]
	\begin{center}
	\includegraphics[width=6.0in]{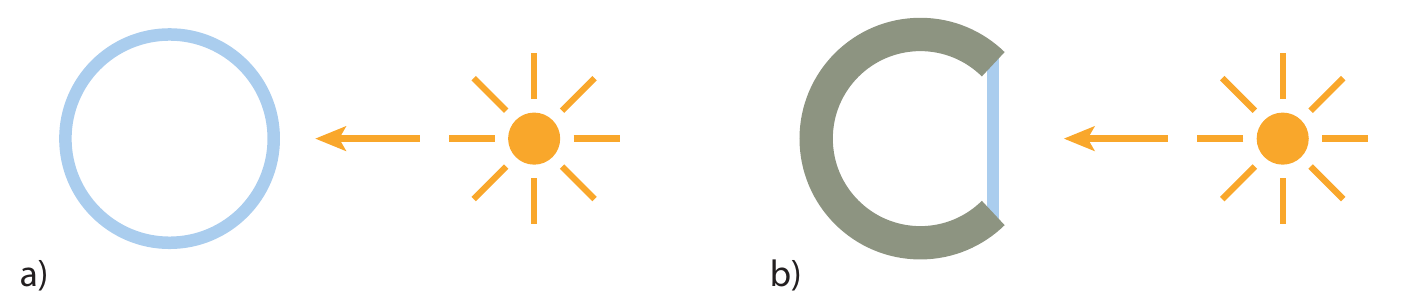}
	\end{center}
	\caption{Geometry for a) spherically symmetric and b) Sun-facing habitat thermal calculations. Blue color represents translucent solid-state greenhouse material of thickness a few cm, while grey represents a thicker layer of opaque, thermally insulating material.}
\label{fig:schematic_2}
\end{figure}

{We estimate the potential for the solid-state greenhouse effect to keep a habitat warm enough for liquid water by considering} the temperature increase from a greenhouse layer with thermal conductivity $\kappa$ and visible extinction coefficient $\gamma_V$. {This} is
\begin{equation}
\Delta T_{max} = \frac{h\alpha F_0 }{\kappa d^2}\label{eq:Tssg},
\end{equation}
where $h$ is the layer thickness  \citep{wordsworth2019enabling}. $\alpha$ can be written as $\E^{-\gamma_V h/ \overline{ \cos \theta}}$, where $\overline {\cos\theta}$ is the average cosine of the solar radiation incident angle relative to the normal vector of the layer surface and $\gamma_V$ is the layer's visible extinction coefficient\footnote{We conservatively neglect the possibility of photons getting absorbed in the greenhouse layer or scattering into the interior when they are removed from the direct beam. We also neglect non-linearities in \eqref{eq:Tssg} due to thermal radiative effects. In the temperature range of interest, this is an acceptable approximation \citep{caps1986infrared}.}. For a sphere, $\overline{\cos \theta} \approx 0.5$ \citep{cronin2014choice}, while for Sun-facing geometry, $\overline{\cos \theta} = 1.0$ (Fig.~\ref{fig:schematic_2}). If $h$ is optimized to maximize warming, $h=h_{max} = \overline{\cos \theta}/\gamma_V$. For a representative value $\gamma_V = 30$~m$^{-1}$, $h_{max}$ is therefore between 1.5 and 3~cm. Combining this result with \eqref{eq:Teq} and \eqref{eq:Tssg} and incorporating geometric effects yields an expression for interior habitat temperature:
\begin{equation}
T_i = T_{eq}  + \Delta T_{max} = \left(\frac{f F_0}{\sigma d^2} \frac {\E^{-1}} {\epsilon} \right)^{1/4} + \frac{f\overline{\cos \theta} F_0\E^{-1}}{\gamma_V\kappa d^2}.\label{eq:SSG}
\end{equation}
This non-linear equation in $d$ can be solved given values for $\alpha$, $\epsilon$ and $\gamma_V$ and a required $T_i$. 

Results are shown in Figure~\ref{fig:temperature} for $T_i = 288$~K ($15^\circ$~C). For the red curve, total spherical symmetry of the habitat is assumed, while for the blue curve, the solid-state greenhouse layer is taken to be Sun-facing, with the rest of the habitat surrounded by a thick layer of opaque insulating material. In such a configuration, the thermal flux from the opaque part of the habitat can be made arbitrarily small as the layer thickness is increased, until the energetic cost of increasing the layer mass becomes prohibitive. Here we take the opaque region thermal loss to be 20\% of the value from the translucent solid-state greenhouse region.\footnote{A systematic derivation indicates that the relevant ratio is $\chi = A_b \kappa_b h_a / A_a \kappa_a h_b$, where the $a$ and $b$ subscripts correspond to the translucent and opaque regions, respectively. For $\chi = 20\%$, $A_b=3A_a$, $\kappa_b=2\kappa_a$ and $h_a = 3$~cm, for example, this would imply $h_b = 90$~cm.}.

As can be seen, maintaining internal temperature at 288~K is possible for a wide range of orbital distances. This calculation assumes a free-floating habitat, but similar considerations apply for habitats {on} the surface of an asteroid, moon {or planet}. Because access to and retention of life-supporting volatiles such as C, N and H becomes a major challenge to habitability close to the Sun (Section~\ref{sec:volatiles}), orbital distances greater than 1~AU ({astronomical unit}) are more favorable. In general, the ability to increase internal habitat temperature is therefore more important than the ability to lower it.

\begin{figure}[h]
	\begin{center}
	\includegraphics[height=3in]{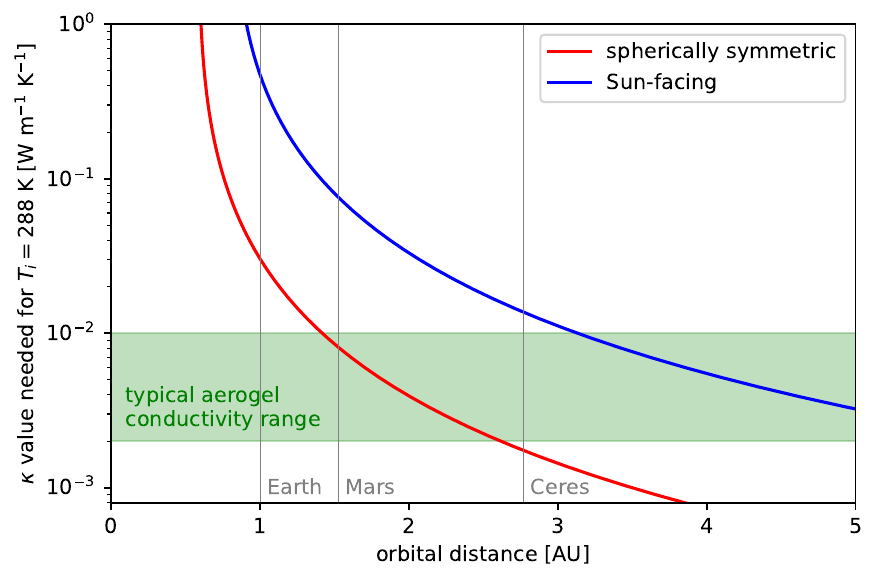}
	\end{center}
	\caption{Passive solid-state warming to enable habitability beyond Earth. Plot shows thermal conductivity of solid-state greenhouse layer vs. orbital distance based on \eqref{eq:SSG},  given a habitat interior temperature of 288~K. Red and blue curves show cases assuming spherically symmetric and Sun-facing geometry, respectively, corresponding to Figs.~\ref{fig:schematic_2}a and b. For these calculations, $\gamma_V = 30$~m$^{-1}$ and $\epsilon = 0.95$. The green shaded area indicates typical thermal conductivities for translucent silica and organic aerogels reported in the literature \citep{dorcheh2008silica,baetens2011aerogel,smalyukh2021thermal}. }
\label{fig:temperature}
\end{figure}

\section{Volatile Loss}\label{sec:volatiles}

The third challenge to extraterrestrial life is volatile loss. All materials have some permeability to atoms and small molecules, and over long timescales the vacuum of space represents an essentially permanent sink for volatile species. In the medium term, volatiles are likely to be a critical resource for any kind of life outside of planetary gravity wells, including humans \citep{schwartz2016near,ellery2020sustainable}. Of course, volatility depends on orbital distance: \ce{H2O} is generally volatile at Earth's orbital distance and closer, but essentially non-volatile beyond a few AU{, as evidenced by the high abundance of surface water ice on small bodies orbiting Jupiter and beyond}. On Mars, which can be considered a semi-benign non-terrestrial environment, volatile loss is important as a local consideration, but not as a global one, given the planet's active water and \ce{CO2} cycles.

To understand the extent of the problem, we can again consider a spherical habitat surrounded by a vacuum. If the habitat has some internal gas pressure $p_i$ for species $i$, the gas flux through the habitat wall
is \citep{baker2012membrane}
\begin{equation}
J \approx D K \frac{p_i}{t},
\end{equation}
where $t$ is the wall thickness and $D$ and $K$ are the diffusion and sorption coefficients for species $i$ in the wall material, respectively. The product $DK$ is usually expressed as the permeability $P$, with SI units of mol~m$^{-1}$~s$^{-1}$~Pa$^{-1}$. The loss rate of gas to space in mol~s$^{-1}$ is $dNdt = -4\pi r^2 J$, where $r$ is the habitat radius. Given that the total number of moles of gas is $N=p_iV/RT$ by the ideal gas law, with $T$ temperature, $V$ volume and $R$ the universal gas constant, the characteristic time for loss of species $i$ in the habitat is 
\begin{equation}
\tau \approx \frac N{|dN/dt|} = \frac{rt}{3 R T P}.
\end{equation}
Permeability for real materials varies widely. To give a everyday example, isoprene rubber has $P$ values of around 150~barrer ($5\times10^{-14}$~mol~m$^{-1}$~s$^{-1}$~Pa$^{-1}$) 
% based on p. 55, Table 2.2 in Baker 
for \ce{CO2} at room temperature \citep{baker2012membrane}. This leads to gas loss timescales from a typical balloon of thickness $t = 10^{-5}$~m and radius $r=0.2$~m of around $\tau\sim 1.5$~hours.

Rubber has a high gas permeability at room temperature because thermal translation and rotation of polymer chains allows for transport of gas molecules through transient microcavities. In glassy polymers, this type of motion is inhibited and diffusion is decreased by 1-2 orders of magnitude. Polylactic acid (PLA), which is a relatively gas-permeable bioplastic, has \ce{O2} permeability of around 1~barrer. Addition of nanocomposites can reduce this value by up to two orders of magnitude by increasing gas molecule diffusion pathways \citep{attallah2021macro}. The gas permeabilities of silica and calcite-based biomaterials do not appear to have been extensively studied, but based on the properties of similar non-biological compounds they would likely be lower still \citep{zouine2007diffusivity}.

Inhibition of volatile escape would be most easily achieved by the same part of the habitat wall responsible for maintaining the pressure differential necessary to stabilize liquid water. For a 1-cm thick layer with permeability $0.1$~barrer, a habitat of radius 10~m and internal temperature 288~K would have $\tau = 13,000$~y. This timescale shows how long it takes to decrease the internal \emph{partial pressure} of a given molecule, so a habitat ecosystem that rapidly transformed volatiles such as \ce{CO2} and \ce{O2} into higher mean molecular mass molecules would inhibit total mass loss further. Additional volatile loss pathways include macroscopic effects due to cracks or defects in the wall material and (hypothetically) use of volatiles for specific purposes, such as propulsion and attitude control in the most autonomous living habitats. 

\section{Radiation, Free Energy and Nutrients}

Other aspects of the space environment present additional obstacles to habitability, although none appear insurmountable. Radiation levels vary widely across the solar system, and at some locations and times are a major hazard to humans \citep{chancellor2014space}, but are unlikely to prevent habitability for microbial life \citep{dartnell2011ionizing}. Shortwave UV solar radiation, especially below 325~nm, is damaging to DNA, proteins and other biomolecules. {However}, it is easily blocked by compounds such as amorphous silica {and reduced iron, which attenuate UV in silicified biofilms and stromatolites today without blocking the visible radiation needed for photosynthesis \citep{cockell2004zones,phoenix2006chilean}.}

Access to sufficient free energy in the form of {visible-wavelength} solar photons is unlikely to be an obstacle in general. Growth of Arctic algae under ice occurs at light levels as low as 0.17~$\mu$mol~photons~m$^{-2}$~s$^{-1}$  \citep[$\sim 0.04$~\Wmsq; ][]{hancke2018extreme}. Rapid growth of many common cyanobacteria and green algae in the laboratory can be achieved at 2-5~\Wmsq. For comparison, the solar flux at the orbit of Jupiter is about 50~\Wmsq. 
% PAR conversion: Set lambda = 580 nm, use E=hc/lambda, umol = 1e-6*NA, get factor of 4.7.
Key nutrients are readily accessible on many asteroids and comets, particularly at orbital distances beyond Earth, where small bodies tend to contain higher abundances of volatile species. Growth of microbial life on a carbonaceous chondrite substrate has recently been demonstrated in a laboratory setting \citep{waajen2022meteorites}. Indeed, in general a range of primordial rocky materials, such as ultramafic rocks or chondritic material in contact with liquid water, will provide biologically accessible CHNOPS elements \citep{cockell2021minimum}. 

Long-term, an additional consideration is the ability of a closed-loop ecosystem to process waste products such as recalcitrant organic matter and to sustain internal redox gradients. On Earth today, most organic material is oxidized in the oceans, but a small amount sinks to the seafloor, where it is eventually thermolyzed in Earth's interior following subduction \citep{berner2004phanerozoic}. Absent such extremes of temperature, a fully closed-loop ecosystem in space would require some internal compartmentalization to establish chemical gradients, and specialist biota capable of breaking down recalcitrant waste products \cite[e.g.,][]{de2016bacterial}. 

\section{Scale and Location}\label{sec:scale}

On Earth, the size of unicellular organisms is limited by factors such as diffusion rates of \ce{O2}, nutrients and {waste products across} the cell membrane (Fig.~\ref{fig:schematic}). The factors that limit the scale of larger organisms are varied and complex, but frequently also emerge from limitations on the chemical transport or diffusion of elements and molecules required for life. The largest relevant scale for terrestrial biota is of course the size of the planet itself. 

For a habitat in an airless environment, the ratio of volume to surface area sets the timescale for loss of volatiles (Section~\ref{sec:volatiles}), which favors larger sizes. However, if photosynthesis is the main source of chemical energy, and liquid \ce{H2O} is the dominant solvent in the habitat interior, the region of productivity will be limited to a euphotic layer of a few tens of meters, beyond which visible light is fully absorbed and photosynthetic primary production is inhibited. The width of this photic zone will depend on how much the habitat wall attenuates visible light, as well as any impurities in the interior that can scatter and/or absorb light. However, as the last few sections have made clear, habitat wall thicknesses on the order of several centimeters are adequate for regulating pressure, temperature and volatile loss, while simultaneously allowing transmission of more than enough visible light for photosynthesis.

For habitats on the surface of a planet or moon, thermal transients are an additional consideration. For example, on the Moon, which has a night duration of $\tau_{night}\sim 14$~days, an uninsulated habitat would require radius {$r \sim 3\tau_{night} \sigma T_{eq}^4/c_p \rho  \Delta T \sim 20$~m} to limit the temperature drop $\Delta T$ to 10~K, given {$\sigma = 5.67\times10^{-8}$~W~m$^{-2}$~K$^{-4}$}, specific heat capacity $c_p \sim 4200$~J~kg$^{-1}$~K$^{-1}$, internal density $\rho \sim 1000$~kg~m$^{-3}$ and external equilibrium temperature $T_{eq}\sim 250$~K. Active or semi-passive thermal control techniques could lower this value, at the cost of additional system complexity.  

Given all these considerations, habitat radii of order 10~m and wall thicknesses of order 0.01~m to 0.1~m are probably close to optimal {for photosynthetic life} (Fig.~\ref{fig:schematic}). In any specific scenario, the exact balance would dependent on wall permeability, internal light extinction rates, physical properties of the wall material, and other factors. Finally, the potential for local wall failure to compromise the habitat might favor redundancy in the form of internal subdivisions. 

Regarding location, two key factors are access to life-essential elements and solar distance. Of these, the latter sets the amount of greenhouse warming required to maintain internal liquid water. These two factors are in tension, as outside of Earth's gravity well, locations where volatiles (especially C, H and N of the biologically important CHNOPS elements) accumulate tend to be those where temperatures are lowest. Locations that are relatively accessible from Earth include polar regions on the Moon and near-Earth volatile-rich asteroids such as 101955 Bennu \citep{hamilton2019evidence}. An intermediate case between Earth and these targets is Mars, which has both an atmospheric source of \ce{CO2} and abundant near-surface \ce{H2O} deposits at high and mid-latitudes \citep{morgan2021availability}. At greater distances, the asteroid belt and Jovian and Saturnian moons possess abundant volatiles, but solar flux is so low that efficient passive thermal control would be needed to maintain habitable temperatures (Fig.~\ref{fig:temperature}).

\begin{figure}[h]
	\begin{center}
	\includegraphics[width=6.5in]{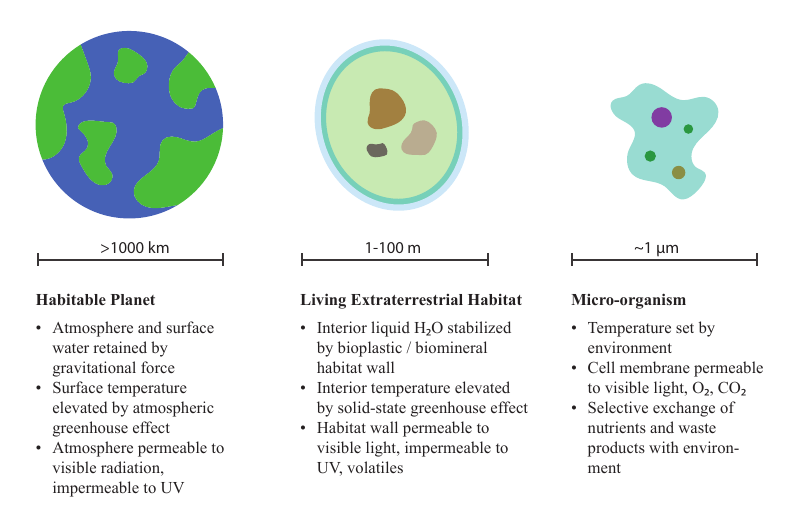}
	\end{center}
	\caption{Schematic summarizing the key similarities and differences between a habitable planet, a minimally complex extraterrestrial habitat, and a micro-organism, with a focus on their physical properties and exchange of mass and energy with their environments.}
\label{fig:schematic}
\end{figure}

\section{Maintenance and Growth}

So far, we have deliberately focused on the basic physical requirements for life in space to keep the discussion as concrete as possible. It should be emphasized that biologically generated habitats for {photosynthetic} life would have many benefits for human {life support} even if they were not completely self-sustaining. Nonetheless, a fully autonomous system capable of regeneration and growth is apparently not prohibited by any physical or chemical constraints, and is therefore interesting to consider a little further.

Regardless of the composition of habitat walls, they will inevitably degrade over time due to UV irradiation, micrometeorite bombardment, and other effects. The timescale on which this occurs would depend on multiple factors, but is likely to be faster than the timescale for volatile loss calculated in Section~\ref{sec:volatiles}. To maintain steady state or achieve growth, regeneration is therefore required. Efficient recycling of old wall material {would} be an essential part of this process. Biological approaches for recycling of bioplastics are already well-developed on Earth, and further advances in synthetic biology will only widen the range of biogenic materials that can be processed 
\citep{rosenboom2022bioplastics}. 

In the least autonomous approach, microbial cultures or plants could be used to create the feedstocks for their own habitat wall material, and then additional processing could occur separately via conventional industrial methods as required. The specific choices of organisms and wall materials in a real application would require detailed study, but amorphous silica, as well as organic polymers such as cellulose, agarose, and lignin, can all be produced by existing photosynthetic life on Earth today.

A more autonomous living habitat would be able to grow its own wall material, just as plant cells regenerate their own walls on the micrometer scale. Wall material could either be created internally and then transported to the outer boundary, or directly synthesized \emph{in situ}. Finally, to grow, a living habitat also requires the ability to absorb new material.  In the most straightforward scenario, nutrients could be processed externally and then added, but a more autonomous habitat would have the ability to absorb new material itself, e.g., through engulfment. This could include the processing of local regolith material and extraction of the required elements.

\section{Silicon and life}\label{sec:NeumannProbes}

Until now, we have focused the discussion on carbon-based life, as this allows a more concrete consideration of requirements. Another type of life that has been proposed to operate in a space environment is a self-replicating `machine'  \citep{freitas1980self,borgue2021near}. In this concept, an autonomous spacecraft capable of propulsion, remote sensing and power generation creates copies of itself, allowing further exploration, potentially without limit. Such a system would be classed as alive according to {many} definitions of the term \citep[e.g., ][]{walker2013algorithmic}.

This topic has been discussed elsewhere, so not too much will be added here. However, it seems clear that the majority of the self-replicating probe ideas are significantly more complex than the living habitats for carbon-based life we have discussed here. Much of this complexity stems from a focus on achieving full self-replicability, as well as the need to rely on semiconductors for power generation and information processing. In the modern semiconductor industry, temperatures of $>1000$~K are required for carbothermal reduction, which is the first step in conversion of \ce{SiO2} to monocrystalline quartz for semiconductor production \citep{lee1977carbothermal}. Additional steps, particularly the microfabrication process, require highly controlled, sterile conditions, which in turn require sophisticated quality control and verification capabilities. 

Interestingly, given how useful silicate-based materials are for control of temperature, pressure and UV radiation (Sections~\ref{sec:pres} and \ref{sec:temp}), carbon-based life is also likely to be dependent on silicon for its survival in an extraterrestrial environment. Although silicon is not an essential element to life on Earth today, many organisms do use silica (\ce{SiO2}) as a structural material, including for spicules within sponges \citep{holzhuter2005silica}, the outer casings of diatoms (frustules) and phytoliths in plants \citep{katz2015silica}. We leave further investigation of the intersection of silicon- and carbon-based life to future work.

\section{Discussion and Conclusions}

In this article, we have argued that non-sentient life is capable of sustaining all the conditions necessary for its own survival in {an} extraterrestrial environment. It is entirely possible that we have underestimated some technical difficulties. Nonetheless, our {calculations indicate} that there are {no obvious physical or chemical limitations} to the survival of self-contained ecosystems beyond Earth, {provided such ecosystems are capable of regulating habitability internally.}

Further research in this area could have benefits for multiple fields. The synthetic biology community has already made a strong case for the use of biotechnology to support humans in space \citep{nangle2020case,berliner2021towards,averesch2023microbial}. {Specific benefits of biotechnology in human life support include medicine and food production, biomaterials for tool and human habitat construction, and fuel for spacecraft propulsion. All of these applications would be aided by biologically generated habitats that could exist in hostile environments with minimal maintenance. } 

The potential benefits to astrobiology of further research on this topic are also high, because it will help to expand thinking beyond the current intense focus on Earth-like environments. The physical principles we consider here, although guided by the requirements of terrestrial biology, are essentially agnostic about the detailed biochemical architecture of life. These principles could be applied with other boundary conditions, such as liquid ammonia as a biochemical solvent, which could lead to new, more general ways to quantify habitability in future. 

For astrobiology, how these alternate forms of life could be detected remotely is a key question. Given that the surfaces of living habitats would have to be partly transparent to allow photosynthesis, one consequence would be a vegetation red edge signature that is observable outside a star system's habitable zone, and potentially outside planetary gravity wells altogether. Detection of such a signature would depend on the number of organisms present and their characteristic size, but standard radiative transfer and remote sensing techniques could be used to estimate detectability thresholds. Living habitats that thermally self-regulate on planets or moons, if sufficiently abundant, could also lead to production of biosignature gases outside of the conventional habitable zone of a given star system.

{Could the kind of biological structures we discuss here evolve naturally, without intelligent intervention? Life on Earth has not yet done this, although it has certainly adapted to an increasingly wide range of environmental conditions over time. Conceivably, a planet or moon with steadily decreasing surface pressure would provide a strong evolutionary drive towards organisms that can regulate temperature and pressure locally. Such a situation could occur due to atmospheric erosion, which is particularly common around red dwarf stars \citep{tarter2007reappraisal}. Investigating the plausibility of different evolutionary pathways for life under alternative planetary boundary conditions will be an interesting topic for future research. }

{More broadly}, the question of how much biology has shaped the habitability of Earth is decades old, and the debate remains active today \citep[e.g.,][]{margulis1974biological,berner2003plants,hilton2020mountains}. Study of the ecology and homeostasis of living habitats in an extraterrestrial setting would bring a new dimension to this debate.  In addition, it could allow {experimental} investigation of whether a biosphere could be maintained by life outside the conventional habitable zone \citep{abbot2011steppenwolf}, given initial planetary conditions that allow life to form \citep{wordsworth2012transient}.
 
Finally, the ideas discussed here can be contrasted with the concept of terraforming. Most frequently proposed for Mars, terraforming is defined as global modification of the climate via industrial technology in order to create conditions suitable for Earth-like life \citep{mckay1991making}. It is controversial and would be phenomenally resource-intensive, as it involves global, irreversible modification of an entire planet. In contrast, biologically generated habitats would pose no more of an environmental concern than any human mission. Indeed, as has been noted elsewhere \citep{santomartino2023toward}, given the effectiveness of biological systems at recycling waste materials, adaptive biology may offer the \emph{most} sustainable and environmentally sound way to support the long-term presence of humans and other complex life beyond Earth in the future.

\section{Code and data availability}

Code to reproduce Figure 3 is available open-source at \emph{https://github.com/wordsworthgroup/space\_habitability}. 

\section{Funding information}

RW acknowledges funding from the Leverhulme Centre for Life in the Universe, Joint Collaborations Research Project Grant G119167, LBAG/312.

\providecommand{\noopsort}[1]{}\providecommand{\singleletter}[1]{#1}%


\begin{thebibliography}{58}
\providecommand{\natexlab}[1]{#1}
\providecommand{\url}[1]{\texttt{#1}}
\expandafter\ifx\csname urlstyle\endcsname\relax
  \providecommand{\doi}[1]{doi: #1}\else
  \providecommand{\doi}{doi: \begingroup \urlstyle{rm}\Url}\fi

\bibitem[Abbot and Switzer(2011)]{abbot2011steppenwolf}
D.~S. Abbot and E.~R. Switzer.
\newblock {The Steppenwolf: A proposal for a habitable planet in interstellar
  space}.
\newblock \emph{The Astrophysical Journal Letters}, 735\penalty0 (2):\penalty0
  L27, 2011.

\bibitem[Arnscheidt et~al.(2019)Arnscheidt, Wordsworth, and
  Ding]{arnscheidt2019atmospheric}
C.~W. Arnscheidt, R.~D. Wordsworth, and F.~Ding.
\newblock Atmospheric evolution on low-gravity waterworlds.
\newblock \emph{The Astrophysical Journal}, 881\penalty0 (1):\penalty0 60,
  2019.

\bibitem[Attallah et~al.(2021)Attallah, Mojicevic, Garcia, Azeem, Chen, Asmawi,
  and Brenan~F.]{attallah2021macro}
O.~A. Attallah, M.~Mojicevic, E.~L. Garcia, M.~Azeem, Y.~Chen, S.~Asmawi, and
  M.~Brenan~F.
\newblock {Macro and micro routes to high performance bioplastics: Bioplastic
  biodegradability and mechanical and barrier properties}.
\newblock \emph{Polymers}, 13\penalty0 (13):\penalty0 2155, 2021.

\bibitem[Averesch et~al.(2023)Averesch, Berliner, Nangle, Zezulka, Vengerova,
  Ho, Casale, Lehner, Snyder, Clark, et~al.]{averesch2023microbial}
N.~J.~H. Averesch, A.~J. Berliner, S.~N. Nangle, S.~Zezulka, G.~L. Vengerova,
  D.~Ho, C.~A. Casale, B.~A.~E. Lehner, J.~E. Snyder, K.~B. Clark, et~al.
\newblock {Microbial biomanufacturing for space-exploration: What to take and
  when to make}.
\newblock \emph{Nature Communications}, 14\penalty0 (1):\penalty0 2311, 2023.

\bibitem[Baetens et~al.(2011)Baetens, Jelle, and Gustavsen]{baetens2011aerogel}
R.~Baetens, B.~P. Jelle, and A.~Gustavsen.
\newblock {Aerogel insulation for building applications: A state-of-the-art
  review}.
\newblock \emph{Energy and Buildings}, 43\penalty0 (4):\penalty0 761--769,
  2011.

\bibitem[Baker(2012)]{baker2012membrane}
R.~A. Baker.
\newblock \emph{{Membrane Transport Theory}}, chapter~2, pages 15--96.
\newblock John Wiley \& Sons, Ltd, 2012.

\bibitem[Berliner et~al.(2021)Berliner, Hilzinger, Abel, McNulty, Makrygiorgos,
  Averesch, Sen~Gupta, Benvenuti, Caddell, Cestellos-Blanco,
  et~al.]{berliner2021towards}
A.~J. Berliner, J.~M. Hilzinger, A.~J. Abel, M.~J. McNulty, G.~Makrygiorgos,
  N.~J.~H. Averesch, S.~Sen~Gupta, A.~Benvenuti, D.~F. Caddell,
  S.~Cestellos-Blanco, et~al.
\newblock {Towards a biomanufactory on Mars}.
\newblock \emph{Frontiers in Astronomy and Space Sciences}, 8:\penalty0 711550,
  2021.

\bibitem[Berner et~al.(2003)Berner, Berner, and Moulton]{berner2003plants}
E.~K. Berner, R.~A. Berner, and K.~L. Moulton.
\newblock {Plants and mineral weathering: Present and past}.
\newblock \emph{{Treatise on Geochemistry}}, 5:\penalty0 605, 2003.

\bibitem[Berner(2004)]{berner2004phanerozoic}
R.~A. Berner.
\newblock \emph{{The Phanerozoic Carbon Cycle: \ce{CO2} and \ce{O2}}}.
\newblock Oxford University Press, 2004.

\bibitem[Borgue and Hein(2021)]{borgue2021near}
O.~Borgue and A.~M. Hein.
\newblock {Near-term self-replicating probes --- A concept design}.
\newblock \emph{Acta Astronautica}, 187:\penalty0 546--556, 2021.

\bibitem[Caps and Fricke(1986)]{caps1986infrared}
R.~Caps and J.~Fricke.
\newblock {Infrared radiative heat transfer in highly transparent silica
  aerogel}.
\newblock \emph{{Solar Energy}}, 36\penalty0 (4):\penalty0 361--364, 1986.

\bibitem[Carr et~al.(1998)Carr, Belton, Chapman, Davies, Geissler, Greenberg,
  McEwen, Tufts, Greeley, Sullivan, et~al.]{carr1998evidence}
M.~H. Carr, M.~J.~S. Belton, C.~R. Chapman, M.~E. Davies, P.~Geissler,
  R.~Greenberg, A.~S. McEwen, B.~R. Tufts, R.~Greeley, R.~Sullivan, et~al.
\newblock {Evidence for a subsurface ocean on Europa}.
\newblock \emph{Nature}, 391\penalty0 (6665):\penalty0 363--365, 1998.

\bibitem[Chancellor et~al.(2014)Chancellor, Scott, and
  Sutton]{chancellor2014space}
J.~C. Chancellor, G.~B.~I. Scott, and J.~P. Sutton.
\newblock {Space radiation: The number one risk to astronaut health beyond low
  earth orbit}.
\newblock \emph{Life}, 4\penalty0 (3):\penalty0 491--510, 2014.

\bibitem[Church(2022)]{church2022picogram}
G.~Church.
\newblock {Picogram-Scale Interstellar Probes via Bioinspired Engineering}.
\newblock \emph{Astrobiology}, 22\penalty0 (12):\penalty0 1452--1458, 2022.

\bibitem[Cockell and Raven(2004)]{cockell2004zones}
C.~S. Cockell and J.~A. Raven.
\newblock {Zones of photosynthetic potential on Mars and the early Earth}.
\newblock \emph{Icarus}, 169\penalty0 (2):\penalty0 300--310, 2004.

\bibitem[Cockell et~al.(2021)Cockell, Wordsworth, Whiteford, and
  Higgins]{cockell2021minimum}
C.~S. Cockell, R.~Wordsworth, N.~Whiteford, and P.~M. Higgins.
\newblock {Minimum units of habitability and their abundance in the Universe}.
\newblock \emph{Astrobiology}, 21\penalty0 (4):\penalty0 481--489, 2021.

\bibitem[Cronin(2014)]{cronin2014choice}
T.~W. Cronin.
\newblock On the choice of average solar zenith angle.
\newblock \emph{Journal of the Atmospheric Sciences}, 71\penalty0 (8):\penalty0
  2994--3003, 2014.

\bibitem[Damant(1937)]{damant1937storage}
G.~C.~C. Damant.
\newblock {Storage of oxygen in the bladders of the seaweed Ascophyllum nodosum
  and their adaptation to hydrostatic pressure}.
\newblock \emph{Journal of Experimental Biology}, 14\penalty0 (2):\penalty0
  198--209, 1937.

\bibitem[Dartnell(2011)]{dartnell2011ionizing}
L.~R. Dartnell.
\newblock Ionizing radiation and life.
\newblock \emph{Astrobiology}, 11\penalty0 (6):\penalty0 551--582, 2011.

\bibitem[De~Gonzalo et~al.(2016)De~Gonzalo, Colpa, Habib, and
  Fraaije]{de2016bacterial}
G.~De~Gonzalo, D.~I. Colpa, M.~H.~M. Habib, and M.~W. Fraaije.
\newblock Bacterial enzymes involved in lignin degradation.
\newblock \emph{Journal of Biotechnology}, 236:\penalty0 110--119, 2016.

\bibitem[Dorcheh and Abbasi(2008)]{dorcheh2008silica}
A.~S. Dorcheh and M.~H. Abbasi.
\newblock Silica aerogel; synthesis, properties and characterization.
\newblock \emph{Journal of materials processing technology}, 199\penalty0
  (1-3):\penalty0 10--26, 2008.

\bibitem[Dyson(1979)]{dyson1979disturbing}
F.~Dyson.
\newblock \emph{{Disturbing the Universe}}.
\newblock Basic Books; Sloan Foundation Science Series, 1979.

\bibitem[Ellery(2020)]{ellery2020sustainable}
A.~Ellery.
\newblock {Sustainable in-situ resource utilization on the Moon}.
\newblock \emph{Planetary and Space Science}, 184:\penalty0 104870, 2020.

\bibitem[Freitas~Jr(1980)]{freitas1980self}
R.~A. Freitas~Jr.
\newblock A self-reproducing interstellar probe.
\newblock \emph{Journal of the British Interplanetary Society}, 33\penalty0
  (7):\penalty0 251--64, 1980.

\bibitem[Goodno and Gere(2020)]{goodno2020mechanics}
B.~J. Goodno and J.~M. Gere.
\newblock \emph{{Mechanics of Materials}}.
\newblock Cengage Learning, 2020.

\bibitem[Hamilton et~al.(2019)Hamilton, Simon, Christensen, Reuter, Clark,
  Barucci, Bowles, Boynton, Brucato, Cloutis, et~al.]{hamilton2019evidence}
V.~E. Hamilton, A.~A. Simon, P.~R. Christensen, D.~C. Reuter, B.~E. Clark,
  M.~A. Barucci, N.~E. Bowles, W.~V. Boynton, J.~R. Brucato, E.~A. Cloutis,
  et~al.
\newblock {Evidence for widespread hydrated minerals on asteroid (101955)
  Bennu}.
\newblock \emph{Nature Astronomy}, 3\penalty0 (4):\penalty0 332--340, 2019.

\bibitem[Hancke et~al.(2018)Hancke, Lund-Hansen, Lamare, H{\o}jlund~Pedersen,
  King, Andersen, and Sorrell]{hancke2018extreme}
K.~Hancke, L.~C. Lund-Hansen, M.~L. Lamare, S.~H{\o}jlund~Pedersen, M.~D. King,
  P.~Andersen, and B.~K. Sorrell.
\newblock {Extreme low light requirement for algae growth underneath sea ice: A
  case study from Station Nord, NE Greenland}.
\newblock \emph{Journal of Geophysical Research: Oceans}, 123\penalty0
  (2):\penalty0 985--1000, 2018.

\bibitem[Hern{\'a}ndez et~al.(2022)Hern{\'a}ndez, Ibarra, Triana,
  Mart{\'\i}nez-Soto, Fa{\'u}ndez, Vasco, Gordillo, Herrera,
  Garc{\'\i}a-Herrera, and Garmulewicz]{hernandez2022agar}
V.~Hern{\'a}ndez, D.~Ibarra, J.~F. Triana, B.~Mart{\'\i}nez-Soto,
  M.~Fa{\'u}ndez, D.~A. Vasco, L.~Gordillo, F.~Herrera, C.~Garc{\'\i}a-Herrera,
  and A.~Garmulewicz.
\newblock {Agar biopolymer films for biodegradable packaging: A reference
  dataset for exploring the limits of mechanical performance}.
\newblock \emph{Materials}, 15\penalty0 (11):\penalty0 3954, 2022.

\bibitem[Hilton and West(2020)]{hilton2020mountains}
R.~G. Hilton and A.~J. West.
\newblock Mountains, erosion and the carbon cycle.
\newblock \emph{Nature Reviews Earth \& Environment}, 1\penalty0 (6):\penalty0
  284--299, 2020.

\bibitem[Holzh{\"u}ter et~al.(2005)Holzh{\"u}ter, Lakshminarayanan, and
  Gerber]{holzhuter2005silica}
G.~Holzh{\"u}ter, K.~Lakshminarayanan, and T.~Gerber.
\newblock {Silica structure in the spicules of the sponge Suberites domuncula}.
\newblock \emph{Analytical and Bioanalytical Chemistry}, 382:\penalty0
  1121--1126, 2005.

\bibitem[Jakosky(2021)]{jakosky2021atmospheric}
B.~M. Jakosky.
\newblock {Atmospheric loss to space and the history of water on Mars}.
\newblock \emph{Annual Review of Earth and Planetary Sciences}, 49:\penalty0
  71--93, 2021.

\bibitem[Kasting et~al.(2014)Kasting, Kopparapu, Ramirez, and
  Harman]{kasting2014remote}
J.~F. Kasting, R.~Kopparapu, R.~M. Ramirez, and C.~E. Harman.
\newblock {Remote life-detection criteria, habitable zone boundaries, and the
  frequency of Earth-like planets around M and late K stars}.
\newblock \emph{Proceedings of the National Academy of Sciences}, 111\penalty0
  (35):\penalty0 12641--12646, 2014.

\bibitem[Katz(2015)]{katz2015silica}
O.~Katz.
\newblock {Silica phytoliths in angiosperms: Phylogeny and early evolutionary
  history}.
\newblock \emph{New Phytologist}, 208\penalty0 (3):\penalty0 642--646, 2015.

\bibitem[Lee et~al.(1977)Lee, Miller, and Cutler]{lee1977carbothermal}
J.~G. Lee, P.~D. Miller, and I.~B. Cutler.
\newblock Carbothermal reduction of silica.
\newblock \emph{Reactivity of Solids}, pages 707--711, 1977.
\newblock Springer, Boston MA.

\bibitem[Li and Ortiz(2014)]{li2014pervasive}
L.~Li and C.~Ortiz.
\newblock Pervasive nanoscale deformation twinning as a catalyst for efficient
  energy dissipation in a bioceramic armour.
\newblock \emph{Nature Materials}, 13\penalty0 (5):\penalty0 501--507, 2014.

\bibitem[Margulis and Lovelock(1974)]{margulis1974biological}
L.~Margulis and J.~E. Lovelock.
\newblock Biological modulation of the earth's atmosphere.
\newblock \emph{Icarus}, 21\penalty0 (4):\penalty0 471--489, 1974.

\bibitem[McKay and Smith(2005)]{mckay2005possibilities}
C.~P. McKay and H.~D. Smith.
\newblock {Possibilities for methanogenic life in liquid methane on the surface
  of Titan}.
\newblock \emph{Icarus}, 178\penalty0 (1):\penalty0 274--276, 2005.

\bibitem[McKay et~al.(1991)McKay, Toon, and Kasting]{mckay1991making}
C.~P. McKay, O.~B. Toon, and J.~F. Kasting.
\newblock {Making Mars habitable}.
\newblock \emph{Nature}, 352\penalty0 (6335):\penalty0 489--496, 1991.

\bibitem[Morgan et~al.(2021)Morgan, Putzig, Perry, Sizemore, Bramson, Petersen,
  Bain, Baker, Mastrogiuseppe, Hoover, et~al.]{morgan2021availability}
G.~A. Morgan, N.~E. Putzig, M.~R. Perry, H.~G. Sizemore, A.~M. Bramson, E.~I.
  Petersen, Z.~M. Bain, D.~M.~H. Baker, M.~Mastrogiuseppe, R.~H. Hoover, et~al.
\newblock {Availability of subsurface water-ice resources in the northern
  mid-latitudes of Mars}.
\newblock \emph{Nature Astronomy}, 5\penalty0 (3):\penalty0 230--236, 2021.

\bibitem[Nangle et~al.(2020)Nangle, Wolfson, Hartsough, Ma, Mason, Merighi,
  Nathan, Silver, Simon, Swett, Thompson, and Ziesack]{nangle2020case}
S.~N. Nangle, M.~Y. Wolfson, L.~Hartsough, N.~J. Ma, C.~E. Mason, M.~Merighi,
  V.~Nathan, P.~A. Silver, M.~Simon, J.~Swett, D.~B. Thompson, and M.~Ziesack.
\newblock {The case for biotech on Mars}.
\newblock \emph{Nature Biotechnology}, 38\penalty0 (4):\penalty0 401--407,
  2020.

\bibitem[Phoenix et~al.(2006)Phoenix, Bennett, Engel, Tyler, and
  Ferris]{phoenix2006chilean}
V.~R. Phoenix, P.~C. Bennett, A.~S. Engel, S.~W. Tyler, and F.~G. Ferris.
\newblock {Chilean high-altitude hot-spring sinters: A model system for UV
  screening mechanisms by early Precambrian cyanobacteria}.
\newblock \emph{Geobiology}, 4\penalty0 (1):\penalty0 15--28, 2006.

\bibitem[Pierrehumbert(2010)]{pierrehumbert2010principles}
R.~T. Pierrehumbert.
\newblock \emph{{Principles of Planetary Climate}}.
\newblock Cambridge University Press, 2010.

\bibitem[Rosenboom et~al.(2022)Rosenboom, Langer, and
  Traverso]{rosenboom2022bioplastics}
J.-G. Rosenboom, R.~Langer, and G.~Traverso.
\newblock Bioplastics for a circular economy.
\newblock \emph{Nature Reviews Materials}, 7\penalty0 (2):\penalty0 117--137,
  2022.

\bibitem[Sagan(1997)]{sagan1997comet}
C.~Sagan.
\newblock \emph{Comet}.
\newblock Ballantine Books, 1997.

\bibitem[Sagan and Salpeter(1976)]{sagan1976particles}
C.~Sagan and E.~E. Salpeter.
\newblock {Particles, environments and possible ecologies in the Jovian
  atmosphere}.
\newblock \emph{NASA Technical Report No. NASA-CR-148170}, 1976.

\bibitem[Santomartino et~al.(2023)Santomartino, Averesch, Bhuiyan, Cockell,
  Colangelo, Gumulya, Lehner, Lopez-Ayala, McMahon, Mohanty,
  et~al.]{santomartino2023toward}
R.~Santomartino, N.~J.~H. Averesch, M.~Bhuiyan, C.~S. Cockell, J.~Colangelo,
  Y.~Gumulya, B.~Lehner, I.~Lopez-Ayala, S.~McMahon, A.~Mohanty, et~al.
\newblock {Toward sustainable space exploration: A roadmap for harnessing the
  power of microorganisms}.
\newblock \emph{Nature Communications}, 14\penalty0 (1):\penalty0 1391, 2023.

\bibitem[Schwartz(2016)]{schwartz2016near}
J.~S.~J. Schwartz.
\newblock {Near-Earth water sources: Ethics and fairness}.
\newblock \emph{Advances in Space Research}, 58\penalty0 (3):\penalty0
  402--407, 2016.

\bibitem[Seager et~al.(2021)Seager, Petkowski, Gao, Bains, Bryan, Ranjan, and
  Greaves]{seager2021venusian}
S.~Seager, J.~J. Petkowski, P.~Gao, W.~Bains, N.~C. Bryan, S.~Ranjan, and
  J.~Greaves.
\newblock {The Venusian lower atmosphere haze as a depot for desiccated
  microbial life: A proposed life cycle for persistence of the Venusian aerial
  biosphere}.
\newblock \emph{Astrobiology}, 21\penalty0 (10):\penalty0 1206--1223, 2021.

\bibitem[Shi et~al.(2015)Shi, Tsai, Camino, Bernard, Yu, and
  Wehner]{shi2015keeping}
N.~N. Shi, C.-C. Tsai, F.~Camino, G.~D. Bernard, N.~Yu, and R.~Wehner.
\newblock {Keeping cool: Enhanced optical reflection and radiative heat
  dissipation in Saharan silver ants}.
\newblock \emph{Science}, 349\penalty0 (6245):\penalty0 298--301, 2015.

\bibitem[Smalyukh(2021)]{smalyukh2021thermal}
I.~I. Smalyukh.
\newblock Thermal management by engineering the alignment of nanocellulose.
\newblock \emph{Advanced Materials}, 33\penalty0 (28):\penalty0 2001228, 2021.

\bibitem[Tarter et~al.(2007)Tarter, Backus, Mancinelli, Aurnou, Backman, Basri,
  Boss, Clarke, Deming, Doyle, et~al.]{tarter2007reappraisal}
J.~C. Tarter, P.~R. Backus, R.~L. Mancinelli, J.~M. Aurnou, D.~E. Backman,
  G.~S. Basri, A.~P. Boss, A.~Clarke, D.~Deming, L.~R. Doyle, et~al.
\newblock {A reappraisal of the habitability of planets around M dwarf stars}.
\newblock \emph{Astrobiology}, 7\penalty0 (1):\penalty0 30--65, 2007.

\bibitem[Tsiolkovsky(1895)]{tsiolkovsky1895speculation}
K.~E. Tsiolkovsky.
\newblock \emph{{Speculation Between Earth and Sky}}.
\newblock Isd-vo AN-SSR, 1895.

\bibitem[Verseux(2020)]{verseux2020bacterial}
C.~Verseux.
\newblock {Bacterial growth at low pressure: A short review}.
\newblock \emph{Frontiers in Astronomy and Space Sciences}, 7:\penalty0 30,
  2020.

\bibitem[Waajen et~al.(2022)Waajen, Prescott, and
  Cockell]{waajen2022meteorites}
A.~C. Waajen, R.~Prescott, and C.~S. Cockell.
\newblock Meteorites as food source on early earth: Growth, selection, and
  inhibition of a microbial community on a carbonaceous chondrite.
\newblock \emph{Astrobiology}, 22\penalty0 (5):\penalty0 495--508, 2022.

\bibitem[Walker and Davies(2013)]{walker2013algorithmic}
S.~I. Walker and P.~C.~W. Davies.
\newblock The algorithmic origins of life.
\newblock \emph{Journal of the Royal Society Interface}, 10\penalty0
  (79):\penalty0 20120869, 2013.

\bibitem[Wordsworth(2012)]{wordsworth2012transient}
R.~Wordsworth.
\newblock Transient conditions for biogenesis on low-mass exoplanets with
  escaping hydrogen atmospheres.
\newblock \emph{Icarus}, 219\penalty0 (1):\penalty0 267--273, 2012.

\bibitem[Wordsworth et~al.(2019)Wordsworth, Kerber, and
  Cockell]{wordsworth2019enabling}
R.~Wordsworth, L.~Kerber, and C.~Cockell.
\newblock {Enabling Martian habitability with silica aerogel via the
  solid-state greenhouse effect}.
\newblock \emph{Nature Astronomy}, 3\penalty0 (10):\penalty0 898--903, 2019.

\bibitem[Zouine et~al.(2007)Zouine, Dersch, Walter, and
  Rauch]{zouine2007diffusivity}
A.~Zouine, O.~Dersch, G.~Walter, and F.~Rauch.
\newblock {Diffusivity and solubility of water in silica glass in the
  temperature range 23--200~$^\circ$C}.
\newblock \emph{Physics and Chemistry of Glasses-European Journal of Glass
  Science and Technology Part B}, 48\penalty0 (2):\penalty0 85--91, 2007.

\end{thebibliography}
\end{document}